\documentclass[conference,hidelinks]{IEEEtran}
\IEEEoverridecommandlockouts
\usepackage{cite}
\usepackage{amsmath,amssymb,amsfonts}
\usepackage{algorithmic}
\usepackage{textcomp}
\usepackage{xcolor}
\usepackage{graphicx}
\usepackage{multirow}
\usepackage{makecell}
\usepackage{csquotes}
\usepackage{framed}
\usepackage{mdframed}
\usepackage{array}
\usepackage[bookmarks=false]{hyperref}

\newcolumntype{L}[1]{>{\raggedright\let\newline\\\arraybackslash\hspace{0pt}}m{#1}}
\newcolumntype{C}[1]{>{\centering\let\newline\\\arraybackslash\hspace{0pt}}m{#1}}
\newcolumntype{R}[1]{>{\raggedleft\let\newline\\\arraybackslash\hspace{0pt}}m{#1}}

\makeatletter
\renewcommand{\@IEEEsectpunct}{\ \,}
\makeatother

\def\BibTeX{{\rm B\kern-.05em{\sc i\kern-.025em b}\kern-.08em
    T\kern-.1667em\lower.7ex\hbox{E}\kern-.125emX}}

\begin{document}

\title{Identification and Remediation of Self-Admitted Technical Debt in Issue Trackers
\thanks{This work was supported by ITEA3 and RVO under grant agreement No. 17038 VISDOM (\url{https://visdom-project.github.io/website}).}
}

\author{Yikun Li, Mohamed Soliman, Paris Avgeriou \\
\IEEEauthorblockA{Bernoulli Institute for Mathematics, Computer Science and Artificial Intelligence \\
University of Groningen \\
Groningen, The Netherlands \\
\{yikun.li, m.a.m.soliman, p.avgeriou\}@rug.nl}
}

\maketitle

\begin{abstract}
Technical debt refers to taking shortcuts to achieve short-term goals, which might negatively influence software maintenance in the long-term.
There is increasing attention on technical debt that is admitted by developers in source code comments (termed as self-admitted technical debt or SATD). But SATD in issue trackers is relatively unexplored.
We performed a case study, where we manually examined 500 issues from two open source projects (i.e. Hadoop and Camel), which contained 152 SATD items. We found that: 1) eight types of technical debt are identified in issues, namely architecture, build, code, defect, design, documentation, requirement, and test debt; 2) developers identify technical debt in issues in three different points in time, and a small part is identified by its creators; 3) the majority of technical debt is paid off, 4) mostly by those who identified it or created it; 5) the median time and average time to repay technical debt are 25.0 and 872.3 hours respectively.
\end{abstract}


\section{Introduction}

Technical debt (TD) refers to taking shortcuts, either deliberately or inadvertently, to achieve short-term goals, which might negatively influence the maintenance and evolution of software in the long term \cite{avgeriou_et_al:DR:2016:6693}. 
Technical debt can be incurred in activities throughout the whole development life cycle, from requirements, to design, implementation, testing, etc. 
There have been several approaches supporting the identification of technical debt in almost all of these activities \cite{alves2016identification}. 
For example, there are approaches detecting code debt by analyzing source code \cite{alves2016identification}, and test debt by analyzing test reports \cite{alves2016identification}. 


A part of technical debt is declared as such by the developers themselves; for example when developers state in source code comments, that something is not right and should be fixed. This has been termed \textit{``Self-Admitted Technical Debt'' (SATD)} \cite{maldonado2015detecting}. 
SATD is often complementary to other types of technical debt items, as it provides information that cannot be uncovered through other means of technical debt identification. 
For example deciding to use a sub-optimal library is likely to be captured in a source code comment but it cannot be detected from source code.
Maldonado and Shihab \cite{maldonado2015detecting} detected five types of SATD (i.e. requirement, code, design, defect, and documentation debt) from source code comments. 


While current work on SATD has focused on source code comments, there are other potentially rich sources of information containing SATD. 
In this paper we focus on SATD in issue trackers, as developers often discuss about technical debt when working on issues. 
There has been some research work exploring technical debt in issue tracking systems \cite{bellomo2016got,dai2017detecting}, showing the possibility of detecting TD through issue trackers, and analyzing the characteristics of technical debt issues, such as opening time, and number of watchers.
However, SATD in issue tracking systems is still relatively unexplored.

The main \textbf{goal} of this paper is to \textit{analyze the types of SATD in issue tracking systems, and to determine how software engineers identify and resolve them}. To achieve our goal, we conducted a case study where we performed a qualitative analysis on a sample of 500 issues.
Specifically, we identified and analyzed sentences in issues that refer to SATD. 
Our findings indicate that: 1) eight types of technical debt are found in issues, namely architecture, build, code, defect, design, documentation, requirement, and test debt; 2) there are three distinct cases of identifying technical debt in issue trackers, while only a small part (13.1\%) of technical debt is identified by its creators; 3) the majority of technical debt is paid off, mostly by those who identified or created it (47.7\% and 44.0\% respectively); 4) the median time and average time spent on technical debt repayment are 25.0 and 872.3 hours.

Our findings provide a number of implications to practitioners and researchers, including: 1) using issue trackers as complementary sources to source code comments for debt detection; 2) developing approaches to detect technical debt, depending on the time that the debt is identified; 3) reporting urgent technical debt in issue trackers, rather than in source code comments, for quicker repayment.

The remainder of this paper is organized as follows. 
In Section~\ref{sec:related}, related work is discussed. Section~\ref{sec:background} presents a typical issue life cycle, accompanied with an example.
The case study design is then elaborated in Section~\ref{sec:approach}, while the results are presented and discussed in Section~\ref{sec:results} and Section~\ref{sec:discussion} respectively.
Finally, threats to validity are evaluated in Section~\ref{sec:validity} and conclusions are drawn in Section~\ref{sec:conclusion}.

\section{Related Work}
\label{sec:related}

In this study, we investigate technical debt in issue trackers, which is a type of SATD.
Thus, we organize the related work into two parts: work related to SATD in general and work related to technical debt in issue trackers. 

\noindent \textbf{Self-admitted Technical debt:}
Potdar and Shihab \cite{potdar2014exploratory} studied self-admitted technical debt in source code comments within four open source projects. 
They found that a range of 2.4\% to 31.0\% of source files contain SATD and 26.3\% to 63.5\% of debt is eventually removed. In a follow-up study, Maldonado and Shihab \cite{maldonado2015detecting} studied five open source projects and discovered the following five types of SATD: design, defect, documentation, requirement, and test debt.

There has also been work related to paying back SATD.
Maldonado \textit{et al.} \cite{maldonado2017empirical} analyzed five Apache projects to study the removal of SATD.
They found that most of SATD is removed by the same person that introduced it, and on median, it takes 18 to 172 days to remove SATD comments.
Zampetti \textit{et al.} \cite{zampetti2018self} also analyzed the removal of SATD in five Java open source projects.
The findings showed that 20\% to 50\% of SATD is removed unintentionally, and 8\% of debt removal is recorded in commit messages.
Our work differs from the work described above, as we look into SATD within issue trackers, instead of source code comments.

\noindent \textbf{Technical debt in issue trackers:}
To the best of our knowledge, only two studies have focused on the detection and comprehension of technical debt in issue trackers. The first, by Bellomo \textit{et al.} \cite{bellomo2016got} presents a classification method for technical debt issues. They manually examined 1,264 issues in four issue trackers from two government projects and two open source industry projects.
From this set, they classified 109 issues as technical debt issues and derived generic characteristics for these issues.
The second study, by Dai and Kruchten \cite{dai2017detecting} analyzed issues from a commercial software issue tracker by reading issue summaries and descriptions.
From 8,149 analyzed issues, they classified 331 as TD issues, and categorized them into six types - defect, requirement, design, code, UI, and architecture debt.
Subsequently, by using machine learning techniques, they trained a classifier with the analyzed issues to automatically classify TD issues.

Our study also classifies issues into types of technical debt (RQ1). But it differs, as it also focuses on how technical debt items are identified (RQ2), and how technical debt items are repaid by developers (RQ3).
Moreover, we analyze issues on the sentence level by reading each sentence in the issue summary, description, and comments.
If a sentence or a group of sentences indicates technical debt, we tag it as a technical debt statement. 
This is different from the aforementioned related studies \cite{bellomo2016got,dai2017detecting} as they both classified whole issues as technical debt issues or non-technical debt issues.
Treating a whole issue as a single type of technical debt may be inaccurate, because software engineers might discuss several types of technical debt in the same issue. For example, in issue HADOOP-6730\footnotemark, software engineers discuss both code debt and test debt. 

\section{Background - Issue Life Cycle}
\label{sec:background}

In general, an issue tracker is a system for issue management.
A managed issue is not only limited to defects but also new features or refactoring.
An issue has its own life cycle, from the time it is created until the time it is resolved. 
The typical steps of this life cycle and an example of each step are shown in Table~\ref{tb:issue_life_cycle}.

\footnotetext{\url{https://jira.apache.org/jira/browse/HADOOP-6730}}

\begin{table}[thpb]
\caption{An example of an issue life cycle.}
\label{tb:issue_life_cycle}
\begin{center}
\resizebox{\columnwidth}{!}{
\def\arraystretch{1.2}
\begin{tabular}{C{0.54cm}|L{0.95cm}|L{3.3cm}|L{3.3cm}}
\hline
\multirow{2}{*}{\textbf{No.}} & \multirow{2}{*}{\textbf{Step}} & \multirow{2}{*}{\textbf{Description}} & \multirow{2}{*}{\textbf{Example} (Hadoop-11074\footnotemark)} \\
 &  &  &  \\
\hline
1 & Create Issue & Usually, software developers create an issue when they find bugs or have new requirements. They first create an issue, which is assigned a unique issue key and describe that issue in detail. & \textit{``Now that hadoop-aws has been created, we should actually move the relevant code into that module, similar to what was done with hadoop-openstack, etc.''} (unique key is Hadoop-11074) \\
\hline
2 & Discuss and Create Patch &  At a later stage, developers start working on it: they comment inside the issue analyzing the problem and sharing their ideas about the solution, and then create a patch to address the issue. & \textit{``HADOOP-11074.patch is attached. This patch does the following: Move the s3 and s3native FS connector code from hadoop- common to hadoop-aws...''} \\
\hline
3 & Code Review & The proposed patch is reviewed by other developers, and feedback is given. If no problem is found in the patch, they proceed to step No.6, otherwise to the next step.  &  \textit{``Can you add an @Ignore on the tests which are failing, so that we can have a green upstream build? +1 once that's addressed.''} \\
\hline
4 & Update Patch &  According to the code review feedback, developers refine the patch and submit it again for another round of code review. &  \textit{``HADOOP-11074.patch.2 is attached. Change the original patch to... This should get Jenkins passing.''} \\
\hline
5 & Code Review & The code is reviewed once more. If it passes, developers proceed to the next step; otherwise, they go back to step No.4. &  \textit{``+1, will commit in an hour or two if there are no more comments.''} \\
\hline
6 & Final Code Commit & The approved patch is committed to the repository with the issue key included in the commit message, and then the issue status is changed to \textit{Resolved}. & \textit{``Patch is committed. Commit message: HADOOP-11074. Move s3-related FS connector code to hadoop-aws.''} \\
\hline
\end{tabular}
}
\end{center}
\vspace{-3mm}
\end{table}
\footnotetext{\url{https://jira.apache.org/jira/browse/HADOOP-11074}}

\section{Case Study Design}
\label{sec:approach}

\begin{figure*}[thpb]
  \centering
  \includegraphics[width=0.88\linewidth]{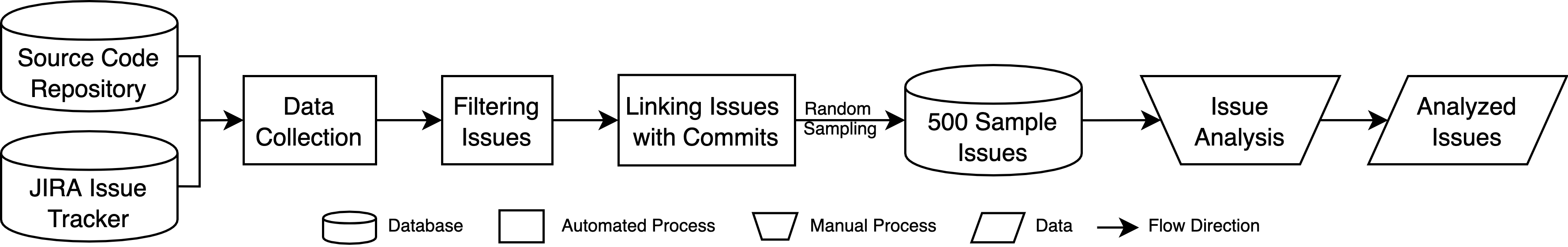}
  \caption{The framework of our approach.}
  \label{f:framework}
  \vspace{-3mm}
\end{figure*}

The goal of this study, formulated according to the Goal-Question-Metric \cite{Solingen:02} template is to ``\textit{\textbf{analyze} issues in issue tracking systems \textbf{for the purpose of} characterizing the technical debt within the issues \textbf{with respect to} the types, the introduction, and the repayment of technical debt \textbf{from the point of view of} software developers \textbf{in the context of} open source software}''.
This goal is refined into three research questions (RQs):

\begin{itemize}
    \item \textbf{(RQ1)} \textit{What types of technical debt are reported in issues?}
    Having knowledge of the types of technical debt could help us understand the strengths and limitations of detecting technical debt in issue trackers.
    For example, we may find that a specific type of technical debt is only detected in issues and not in other sources, or that it is mostly detected in issues. 
    That can help in proposing approaches for detecting technical debt that combine different sources. Although Dai and Kructhen \cite{dai2017detecting} also studied types of debt in issues, they only analyzed the issue summary and description.
    In contrast, we analyze entire issues (including the comments) at the level of sentences.
    
    \item \textbf{(RQ2)} \textit{When do software developers identify technical debt in issues?}
    This RQ aims at understanding the point in time that debt is identified in issue trackers. For example, technical debt can be incurred when working on an issue, or it can exist beforehand and the issue is created to address it.
    This can help researchers to tune their TD detection approaches depending on when it is identified.
    For example, if the technical debt is added to a patch and eventually the patch is rejected (not committed), the debt is not added to the system. 
    In this case, an approach may falsely detect this debt item in a code review statement regarding that (rejected) patch.
    
    \item \textbf{(RQ3)} \textit{How do software engineers resolve technical debt in issues?} 
    This is further refined into 3 sub-questions:
    
    \begin{itemize}
        \item \textbf{(RQ3.1)} \textit{How much technical debt is resolved?}
        Quantifying how much technical debt is paid off, helps us understand developers' attitudes towards technical debt and of course the magnitude of the problem.
        For instance, if most of the debt is discussed and resolved, it would imply that developers are aware of the harmfulness of technical debt and take action resolving it. It would also imply that technical debt in issues does not pose a critical threat.
    
        \item \textbf{(RQ3.2)} \textit{Who resolves technical debt?}
        Technical debt can be resolved by those who created it, those who discovered it, or by others.
        This aids in understanding the practices of developers, e.g. if those that incur debt take the responsibility to resolve it. It can also be used to assist with debt repayment; for example if the debt creator did not resolve it, another developer may need more documentation to understand the problem well enough in order to solve it.
        
        \item \textbf{(RQ3.3)} \textit{How long does it take to resolve technical debt?}
        Knowing how long it normally takes to repay technical debt after discovering it, is helpful for technical debt management.
        Technical debt that is long-lived causes extra maintenance effort and should thus be prioritized for remediation.
    \end{itemize}
\end{itemize}

Fig.~\ref{f:framework} shows the approach we follow to answer the research questions.
The four individual processes (automated and manual) are explained in the following sub-sections.

\subsection{Data collection}

To answer the research questions, we looked into Apache Java projects since they are of high quality and supported by mature communities.
To select Apache projects pertinent to our study goal, we set the following criteria:

\begin{enumerate}
    \item Both the issue tracking project and the source code repository are publicly available and well-maintained.
    
    \item They have at least 1,000,000 source lines of code (SLOC) and 10,000 issues in the issue tracker. This is to ensure sufficient complexity.
    
    \item Source code commits involve their associated issue keys within their comments. 
    This is important to support linking commits (in the source code repository) with issues (in the issue tracker). 
    This is further motivated in Section~\ref{sec:link}.
    
    \item They are commonly used in other SATD studies (e.g. \cite{maldonado2017empirical}).
    This allows us to compare the results between our study and other SATD studies.
\end{enumerate}

Based on these criteria, we selected Hadoop\footnote{\url{https://hadoop.apache.org}} and Camel\footnote{\url{https://camle.apache.org}}. 
Both projects were studied for SATD \cite{maldonado2017empirical}, were developed in Java, used Git as a source code repository and JIRA\footnote{\url{https://jira.apache.org}} as an issue tracker.
We analyzed the latest released versions on Jan 16, 2020.
Table~\ref{tb:projects} shows some details for the two projects. 
The number of Java files and SLOC are calculated using the LOC tool\footnote{\url{https://github.com/cgag/loc}}.
The number of contributors is obtained from GitHub.
We used the JIRA Python package to extract all Hadoop and Camel issues from the online server and stored them in a local database; then we counted the number of issues.

\begin{table}[th]
\caption{Details of chosen projects.}
\label{tb:projects}
\begin{center}
\resizebox{\columnwidth}{!}{
\def\arraystretch{1.2}
\begin{tabular}{c|c|c|c|c|c}
\hline
\multirow{2}{*}{\textbf{Project}} & \multirow{2}{*}{\textbf{\# Java files}} & \multirow{2}{*}{\textbf{SLOC}} & \multirow{2}{*}{\textbf{\# Contributors}} & \multirow{2}{*}{\textbf{\# Issues}} & \multirow{2}{*}{\textbf{\# Filtered issues}} \\
 & & & & \\
\hline
Hadoop & 10,918 & 1,700,501 & 259 & 16,808 & 6,685 \\
\hline
Camel & 17,585 & 1,196,790 & 583 & 14,411 & 12,259 \\
\hline
\end{tabular}
}
\end{center}
\end{table}

\begin{table*}[thpb]
\caption{Definitions of indicators of different types of technical debt in issue trackers.}
\label{tb:our_indicators}
\begin{center}
\resizebox{2\columnwidth}{!}{
\def\arraystretch{1.2}
\begin{tabular}{L{2.5cm}|L{3.7cm}|C{0.9cm}|L{8.8cm}}
\hline
\textbf{Type} & \textbf{Indicator} & \textbf{Reused} & \textbf{Definition} \\
\hline
Architecture debt & Violation of modularity & $\bullet$ &  Because shortcuts were taken, multiple modules became inter-dependent, while they should be independent. \\
\cline{2-4}
 &  Using obsolete technology & $\circ$ & Architecturally-significant technology has become obsolete. \\
\hline
Build debt & Under- or over-declared dependencies & $\bullet$ & Under-declared dependencies: dependencies in upstream libraries are not declared and rely on dependencies in lower level libraries. \newline Over-declared dependencies: unneeded dependencies are declared. \\
\cline{2-4}
 & Poor deployment practice & $\circ$ & The quality of deployment is low that compile flags or build targets are not well organized. \\
\hline
Code debt & Complex code & $\circ$ & Code has accidental complexity and requires extra refactoring action to reduce this complexity. \\
\cline{2-4}
 & Dead code & $\circ$ & Code is no longer used and needs to be removed. \\
\cline{2-4}
 & Duplicated code & $\bullet$ & Code that occurs more than once instead of as a single reusable function. \\
\cline{2-4}
 & Low-quality code & $\circ$ & Code quality is low, for example because it is unreadable, inconsistent, or violating coding conventions. 
 \\
\cline{2-4}
 & Multi-thread correctness & $\bullet$ & Thread-safe code is not correct and may potentially result in synchronization problems or efficiency problems. \\
\cline{2-4}
 & Slow algorithm & $\bullet$ & A non-optimal algorithm is utilized that runs slowly. \\
\hline
Defect debt & Uncorrected known defects & $\bullet$ & Defects are found by developers but ignored or deferred to be fixed. \\
\hline
Design debt & Non-optimal decisions & $\circ$ & Non-optimal design decisions are adopted. \\
\hline
Documentation debt & Outdated documentation & $\bullet$ & A function or class is added, removed, or modified in the system, but the documentation has not been updated to reflect the change. \\
\cline{2-4}
 & Low-quality documentation & $\circ$ & The documentation has been updated reflecting the changes in the system, but quality of updated documentation is low. \\
\hline
Requirement debt & Requirements partially implemented & $\circ$ & Requirements are implemented, but some are not fully implemented. \\
\cline{2-4}
 & Non-functional requirements not fully satisfied & $\circ$ & Non-functional requirements (e.g. availability, capacity, concurrency, extensibility), as described by scenarios, are not fully satisfied. \\
\hline
Test debt & Expensive tests & $\circ$ & Tests are expensive, resulting in slowing down testing activities. Extra refactoring actions are needed to simplify tests.  \\
\cline{2-4}
 & Lack of tests & $\circ$ & A function is added, but no tests are added to cover the new function. \\
\cline{2-4}
 & Low coverage & $\bullet$ & Only part of the source code is executed during testing. \\
\hline
\end{tabular}
}
\end{center}
\end{table*}

\subsection{Filtering issues}
\label{sec:filter}

To ensure that we study issues with a complete life cycle (as shown in Table~\ref{tb:issue_life_cycle}), we applied two filtering criteria:

\begin{enumerate}
    \item \textbf{Issue status:} 
    Since we are aiming at studying technical debt items that were resolved, we focus on issues that are done or closed.
    Thus, we removed all issues with status \textit{Open} or \textit{Pending Closed}.
    
    \item \textbf{Availability of issue key in commits:}
    Although some issues have their status set to \textit{Resolved} and developers commented that the patches are successfully committed to the repositories, we cannot find the related commits in Git. This is mostly because developers did not include the issue key in the corresponding commit messages.
    We also exclude these issues, since we need the commit information to be able to answer RQ3 on debt repayment.
\end{enumerate}

The final number of issues after filtering is listed in the rightmost column of Table~\ref{tb:projects}.

\subsection{Linking issues with commits}
\label{sec:link}

In order to determine how software engineers actually resolve technical debt (i.e. answering RQ3), we have to capture the code commits associated with an issue. 
This information is needed to determine the software developers responsible for repaying technical debt (RQ3.2) and the time for this repayment (RQ3.3).

Since in the previous step, we ensured that the commit messages contain the related issue keys, we use those keys to link issues with commits.
In practice, we first output the Git commit log, and match the issue key by applying a regular expression to the commit log.
Then all matched commits (including commit date, commit message, and commit author) are inserted into the issue holding the issue key ordered by time, and then the issue with commit information is stored in a local database.

\subsection{Issue manual analysis}

The filtering step resulted in 18944 issues that fulfill our criteria (see Section~\ref{sec:filter}): 6685 for Hadoop and 12259 for Camel. 
Since manually analyzing issues is extremely time-consuming, we are only able to analyze a subset.
From this set, we randomly selected a sample of 500 issues for analysis: 250 issues from each project (i.e. Hadoop and Camel). The size of our sample is in line with similar studies, e.g. Zaman \textit{et al.} analyzed 400 issues to study performance bugs \cite{zaman2012qualitative}.
To analyze issues for technical debt, we followed the instructions for qualitative analysis proposed by Runeson \textit{et al.} \cite{runeson2012case}.  
We used a professional qualitative content analysis tool (ATLAS.ti\footnote{\url{https://atlasti.com}}) to annotate relevant sentences within the sample issues.

To answer RQ1, we performed a classification using an existing framework from Alves \textit{et al.} \cite{alves2014towards}. 
This framework provides basic types of technical debt, with high-level definitions and a list of indicators per type.
Using these types, we annotated sentences within issues, referring to existing debt or resolving debt. 
We read each sentence in issue summary, description, and comments. If a sentence or a group of sentences indicated a certain type of technical debt, we tagged it with that type and relevant indicators.

The issues were independently annotated by the first and second author. 
The differences between the two authors supported refining the types and indicators of technical debt from the original framework of Alves \textit{et al.} \cite{alves2014towards}.
For example, we added the indicator \textit{Requirements Partially Implemented} to the requirement debt type.
The refined classification framework that resulted from this step is presented in Table~\ref{tb:our_indicators}.
The \textit{Reused} column refers to whether the indicators are reused directly from the study of Alves \textit{et al.} (``$\bullet$'' symbol) or they were created inductively during the qualitative analysis (``$\circ$'' symbol).
The original framework of Alves \textit{et al.}, can be found in the replication package\footnote{\url{http://www.cs.rug.nl/search/uploads/Resources/li\_soliman\_avgeriou\_seaa2020.zip}}. 
The classification resulted in 152 annotated statements with different technical debt types and indicators, which are also available in the replication package.

To mitigate the risk of bias, we evaluated the level of agreement between the classifications of the two authors using Cohen's kappa coefficient \cite{fleiss1981measurement}; this is commonly used to measure inter-rater reliability. 
The calculated level of agreement between the two authors is 0.757 based on a sample consisting of 15\% of all technical debt statements, which is considered excellent according to the work of Fleiss \textit{et al.} \cite{fleiss1981measurement}.

Next, we revisited all identified technical debt to obtain information to answer RQ2 and RQ3.
More specifically, for RQ2, we annotated text with information regarding the identification of technical debt items within the issue life cycle. 
Regarding RQ3, for each technical debt item, we read the related issue comments and the corresponding commit messages (see Section~\ref{sec:link}) to identify information on debt remediation. 
If indeed there was such information, we noted it down, as well as the person who resolved the item and the time between reporting and resolving it.

\section{Results}
\label{sec:results}

\subsection{(RQ1) What types of technical debt are reported?}
\label{sec:rq1}

We found eight types of technical debt in issues: architecture, build, code, defect, design, documentation, requirement, and test debt. 
For each type we found one or more indicators. 
In the following paragraphs, we report on the associated indicators for each type, also providing a quote from actual issues to exemplify each indicator.

\noindent\textbf{Architecture debt:}
problems that are architecturally significant, i.e. they are hard to change.
Most of the debt in this type relates to the indicator \textit{Violation of Modularity}.

\begin{displayquote}
\textit{``It would be good if these were moved into their own module...''} - [Camel-4543]

\end{displayquote}

Some architecture debt is caused by \textit{Using Obsolete Technology}.

\begin{displayquote}
\textit{``The camel-atom component is using an ancient incubator version of abdera which will make it hard to work with camel-cxf.''} - [Camel-4132]
\end{displayquote}

\noindent\textbf{Build debt:} 
issues that make building (i.e. source code compilation to artifacts) harder or more time-consuming.
Most of the identified build debt is caused by \textit{Over- or Under-Declared Dependencies}.

\begin{displayquote}
\textit{``Avoid the redundant direct dependency on log4j by the components.''} - [Camel-4331]

\textit{``Compiling for Fedora revels a missing declaration for javax.annotation.Nullable. This is the result of a missing explicit dependency on...''} - [Hadoop-10067]
\end{displayquote}

The rest of build debt is caused by \textit{Poor Deployment Practice}.

\begin{displayquote}
\textit{``Rationalize the way architecture-specific sub-components are built with ant in branch-1. This is a matter of maintainability and understandability, and therefore robustness under future changes in build.xml.''} - [Hadoop-8364]
\end{displayquote}

\noindent\textbf{Code debt:}
issues in source code, which negatively influence the maintenance of software.
Most of the code debt is caused by \textit{Low-Quality Code}.

\begin{displayquote}

\textit{``This will lead to very unmaintainable code. We absolutely do not want to have nested retries for different contexts.''} - [Hadoop-3198]

\end{displayquote}

A few code debt items result from \textit{Slow Algorithm}.

\begin{displayquote}
\textit{``\#query() does O(N) calls LinkedList\#get() in a loop, rather than using an iterator. This makes query O(N\textasciicircum 2), rather than O(N).''} - [Hadoop-8866]
\end{displayquote}

\textit{Multi-Thread Correctness} is another factor causing code debt.

\begin{displayquote}
\textit{``EnsureInitialized() forced many frequently called methods to unconditionally acquire the class lock.''} - [Hadoop-9748]

\end{displayquote}

The rest of the code debt is caused by \textit{Dead Code}, \textit{Duplicated Code}, and \textit{Complex Code}.

\begin{displayquote}
\textit{``As we don't use the CxfSoap component any more, it's time to clean it up.''} - [Camel-2535]

\textit{``I am concerned about the code duplication this brings.''} - [Hadoop-6381]


\textit{``...can be simplified to the following so there aren't so many return statements to track.''} - [Hadoop-10169]
\end{displayquote}

\noindent\textbf{Defect debt:}
known defects that are deferred to be fixed.
All defect debt items are caused by \textit{Uncorrected Known Defects}.

\begin{displayquote}
\textit{``This works in 2.12.x onwards. Hunting this down on 2.11.x is low priority. End users is encourage to upgrade if they really need this.''} - [Camel-6735]
\end{displayquote}

\noindent\textbf{Design debt:}
shortcuts or non-optimal decisions taken in detailed design.
All design debt results from \textit{Non-Optimal Decisions}.

\begin{displayquote}
\textit{``Instead of passing a long[] you should pass a struct that implements Writable.''} - [Hadoop-481]

\textit{`Extending the Trash API might be ok in the short term but does not sound too appealing from a long-term perspective.''} - [Hadoop-2815]
\end{displayquote}

\noindent\textbf{Documentation debt:}
when the software is modified, the documentation is not updated to reflect the changes or the quality of updated documentation is low.
Most of this type of debt is caused by \textit{Outdated Documentation}.

\begin{displayquote}
\textit{``The maven reports is just getting to old and intermixed with 1.x and trunk releases.''} - [Camel-1846]

\end{displayquote}

The second indicator is \textit{Low-Quality Documentation}.

\begin{displayquote}
\textit{``I agree to improve documentation to make it clear that...''} - [Hadoop-12672]

\end{displayquote}

\noindent\textbf{Requirement debt:} 
when the requirements specification is not in line with the actual implementation.
Some requirement debt is caused by \textit{Requirements Partially Implemented}.

\begin{displayquote}
\textit{``The only feature which we don't support is correlated message groups. That requires a bit more work and also may complicated...''} - [Camel-1669]

\end{displayquote}

Another common cause concerns \textit{Non-Functional Requirements Not Being Fully Satisfied}. 
In the example below, concurrency is not fully satisfied.

\begin{displayquote}
\textit{``Definition requires the implementations for its interfaces should be thread-safe. HarFsInputStream doesn't implement these interfaces with tread-safe, this JIRA is to fix this.''} - [Camel-5587]
\end{displayquote}

\noindent\textbf{Test debt:} 
shortcuts or non-optimal decisions taken in testing that negatively affect maintainability.
Most test debt is caused by \textit{Lack of Tests}.

\begin{displayquote}
\textit{``There are no XQuery specific tests.''} - [Camel-201]
\end{displayquote}

The other major cause of test debt is \textit{Low Coverage}.

\begin{displayquote}
\textit{``Some of the test code doesn't check for correct error codes to correspond with the wrapped exception type.''} - [Hadoop-11103]
\end{displayquote}

Finally, some test debt results from \textit{Expensive Tests}.

\begin{displayquote}
\textit{``I see recent hadoop-hdfs test runs have been taking 2.5 hours. This one (new patch) was 45 minutes.''} - [Hadoop-11670]
\end{displayquote}

\begin{table}[thpb]
\caption{Types and indicators of technical debt.}
\label{tb:statistics_type_projects}
\begin{center}
\resizebox{\columnwidth}{!}{
\def\arraystretch{1.2}
\begin{tabular}{c|c|c|c|c}
\hline
\textbf{Type} & \textbf{Indicator} & \textbf{\#}\footnotemark & \textbf{\#}\footnotemark[\value{footnote}] & \textbf{\%} \\
\hline
\multirow{2}{*}{\makecell{Architecture \\debt}} & Violation of modularity & 8 & \multirow{2}{*}{10} & \multirow{2}{*}{6.6} \\
\cline{2-3}
 & Using obsolete technology & 2 &  \\
\hline
\multirow{2}{*}{\makecell{Build debt}} & Over- or under-declared dependencies & 5 & \multirow{2}{*}{6} & \multirow{2}{*}{3.9} \\
\cline{2-3}
 & Poor deployment practice & 1 &  \\
\hline
\multirow{6}{*}{\makecell{Code debt}} & Complex code & 2 & \multirow{6}{*}{59} & \multirow{6}{*}{38.8} \\
\cline{2-3}
 & Dead code & 12 &  \\
\cline{2-3}
 & Duplicated code & 6 &  \\
\cline{2-3}
 & Low-quality code & 36 &  \\
\cline{2-3}
 & Multi-thread correctness & 1 &  \\
\cline{2-3}
 & Slow algorithm & 2 &  \\
\hline
\multirow{1}{*}{\makecell{Defect debt}} & Uncorrected known defects & 4 & 4 & 2.6 \\
\hline
\multirow{1}{*}{\makecell{Design debt}} & Non-optimal decisions & 8 & 8 & 5.3 \\
\hline
\multirow{2}{*}{\makecell{Documentation \\debt}} & Low-quality documentation & 16 & \multirow{2}{*}{33} & \multirow{2}{*}{21.7} \\
\cline{2-3}
 & Outdated documentation & 17 &  \\
\hline
\multirow{3}{*}{\makecell{Requirement \\debt}} & Requirements partially implemented & 3 & \multirow{3}{*}{4} & \multirow{3}{*}{2.6} \\
\cline{2-3}
 & \makecell{Non-functional requirements not \\being fully satisfied} & 1 &  \\
\hline
\multirow{3}{*}{\makecell{Test debt}} & Expensive tests & 1 & \multirow{3}{*}{28} & \multirow{3}{*}{18.4} \\
\cline{2-3}
 & Lack of tests & 20 &  \\
\cline{2-3}
 & Low coverage & 7 &  \\
\hline
\end{tabular}
}
\end{center}
\vspace{-5mm}
\end{table}
\footnotetext{The symbol \# refers to the number of instances.}

Table~\ref{tb:statistics_type_projects} presents an overview of technical debt types and indicators in the examined issues.
We observe that code, documentation, and test debt are the three most common types (with 38.8\%, 21.7\%, and 18.4\% respectively).
Furthermore, the three most common indicators are \textit{Low-quality Code}, \textit{Lack of Tests}, and \textit{Outdated Documentation}.

Finally, since we annotated technical debt on the sentence level (instead of the issue level), an issue may contain more than one types of technical debt.
Table~\ref{tb:td_concurrence} presents how many issues contain zero, one or more types of technical debt in issues.
As we can see, 24 out of 117 issues (20\%) that contain technical debt, contain more than one type. 
This validates our choice to analyze issues at the level of sentences; if we had performed the analysis at the level of issues, we would have missed the additional technical debt types per issue.

\begin{table}[th]
\caption{Numbers of types of technical debt in issues.}
\label{tb:td_concurrence}
\begin{center}
\resizebox{\columnwidth}{!}{
\def\arraystretch{1.2}
\begin{tabular}{C{5cm}|C{1.25cm}|C{1.25cm}}
\hline
\textbf{Issue description} & \textbf{\# Issues} & \textbf{\% Issues} \\
\hline
Does not contain technical debt & 383 & 76.6 \\
\hline
Contains one type of technical debt & 93 & 18.6 \\
\hline
Contains two types of technical debt & 21 & 4.2 \\
\hline
Contains three types of technical debt & 2 & 0.4 \\
\hline
Contains four types of technical debt & 1 & 0.2 \\
\hline
\end{tabular}
}
\end{center}
\vspace{-5mm}
\end{table}

\begin{framed} 
\noindent \textit{Eight types of technical debt are found in issue trackers: architecture, build, code, defect, design, documentation, requirement, and test debt.
The three most common types are code, documentation, and test debt (i.e. 38.8\%, 21.7\%, and 18.4\%). About one fifth of the issues that contain technical debt, contain more than one type.
}
\end{framed}

\subsection{(RQ2) When do software engineers identify technical debt?}
\label{sec:rq2}

We observed three distinct cases of technical debt being identified in issue trackers:

\begin{enumerate}
    \item \textit{Identifying technical debt before creating an issue (i.e. debt is the reason for creating the issue):} When developers spot an existing technical debt item in the system, they report it in an issue tracker to be resolved. For instance, a developer found low-quality code, which complicates debugging; thus, he/she created a new issue: 
    
    \begin{displayquote}
    \textit{``If the user doesn't setup the right camel context for the context component. The exception we got is misleading, we need to throw more meaningful exception for it.''} - [Camel-5714]
    \end{displayquote}
    
    \item \textit{Identifying technical debt during code review:} As explained in Section~\ref{sec:background}, software engineers perform code reviews by creating and reviewing code patches in issue trackers. When a code reviewer identifies technical debt items in a code patch, he/she discusses it with other developers to determine, if the identified technical debt should be resolved or committed to the system.
    For example, during a code review, a developer found that a shortcut was taken. Thus, he/she commented on a patch:
    
    \begin{displayquote}
    \textit{``The patch looks good to me... It would be better if we can add an upper limit for the size of the GSet.''} - [Hadoop-9763]
    \end{displayquote}
    
    \item \textit{Identifying technical debt after a patch is committed:} Technical debt can exist in a patch but go undetected through the code review; after the patch is committed, a developer may notice the debt in the commit and report it.
    For instance, after a command patch is committed to the repository, a developer noticed that documentation is not updated accordingly:
    
    \begin{displayquote}
    \textit{``We need to update the documentation with the new command.''} - [Camel-8101]
    \end{displayquote}
\end{enumerate}

\begin{table}[th]
\caption{Technical debt identification cases.}
\label{tb:td_introduction}
\begin{center}
\resizebox{\columnwidth}{!}{
\def\arraystretch{1.2}
\begin{tabular}{C{1cm}|C{1.5cm}|C{0.5cm}|C{0.5cm}|C{0.5cm}|C{0.5cm}|C{0.5cm}|C{0.5cm}}
\hline
\multirow{2}{*}{\textbf{Project}} & \multirow{2}{*}{\textbf{\# Identified}} & \multicolumn{2}{c|}{\textbf{Case 1}} & \multicolumn{2}{c|}{\textbf{Case 2}} & \multicolumn{2}{c}{\textbf{Case 3}} \\
\cline{3-8}
 & & \textbf{\#\footnotemark[\value{footnote}]} & \textbf{\%} & \textbf{\#\footnotemark[\value{footnote}]} & \textbf{\%} & \textbf{\#\footnotemark[\value{footnote}]} & \textbf{\%} \\
\hline
Hadoop & 101 & 41 & 40.6 & 57 & 56.4 & 3 & 3.0 \\
\hline
Camel & 51 & 27 & 52.9 & 13 & 25.5 & 11 & 21.6 \\
\hline
Total & 152 & 68 & 44.7 & 70 & 46.1 & 14 & 9.2 \\
\hline
\end{tabular}
}
\end{center}
\vspace{-2mm}
\end{table}

To gain a better understanding of how technical debt is identified, Table~\ref{tb:td_introduction} presents the count of technical debt items for the three aforementioned cases.
Clearly, the first and second cases represent the majority (44.7\% and 46.1\% respectively) in these projects.
Compared with Camel, there is 30.9\% more debt introduced in Hadoop with the second case and 18.6\% less debt introduced with the third case.
This means that more technical debt is identified during code reviews (on patches) than after the patch is committed in the system in Hadoop compared with Camel.

\begin{table}[th]
\caption{Technical debt reporters.}
\label{tb:td_creator_info}
\begin{center}
\resizebox{0.92\columnwidth}{!}{
\def\arraystretch{1.2}
\begin{tabular}{C{1cm}|C{1.25cm}|C{1.25cm}|C{1.25cm}|C{1.25cm}}
\hline
\multirow{2}{*}{\textbf{Project}} & \multicolumn{2}{c|}{\textbf{Reported by creators}} & \multicolumn{2}{c}{\textbf{Reported by others}} \\
\cline{2-5}
 & \textbf{\#\footnotemark[\value{footnote}]} & \textbf{\%} & \textbf{\#\footnotemark[\value{footnote}]} & \textbf{\%} \\
\hline
Hadoop & 4 & 6.7 & 56 & 93.3 \\
\hline
Camel & 7 & 29.2 & 17 & 70.8 \\
\hline
Total & 11 & 13.1 & 73 & 86.9 \\
\hline
\end{tabular}
}
\end{center}
\vspace{-2mm}
\end{table}

Moreover, we also investigate who reported the debt: the developers who created it in the first place or those who discovered it.
Since technical debt identified in the first case already exists in the system, information on who created it is not contained in issue trackers; thus, such information is obtained only for technical debt identified in the second and third cases.
Table~\ref{tb:td_creator_info} presents an overview on who reported the technical debt. 
We find that on average most of the debt is reported by other developers (i.e. 86.9\%), and a small part is self-reported (reported by those that created it).
Camel has a higher percentage of self-reported debt than Hadoop, but the vast majority of its debt is still reported by others (i.e. 70.8\% versus 29.2\%).
This may mean that most developers create technical debt unintentionally.

\begin{framed} 
\noindent \textit{There are three cases of identifying technical debt in issue trackers: discovering existing debt and creating an issue for it, identifying debt in a patch during code review, or after the patch is committed in the system.
Most of the technical debt is identified in the first and second cases.
A small part of the debt is reported by its creators, while most is reported by other developers.
}
\end{framed}

\subsection{(RQ3) How do software engineers resolve technical debt?}
\label{sec:rq3}

\subsubsection{\textbf{(RQ3.1) How much technical debt is paid off?}}
\label{sec:repaid_rate}

Table~\ref{tb:repayment_statistics} presents the amounts and percentages of technical debt items that are identified and resolved.
We can see that most of the identified technical debt is actually resolved in both Hadoop and Camel (i.e. 71.3\% and 72.5\%, respectively). 
This indicates that, when technical debt is reported in issue trackers, it will likely be resolved. 
In other words, most software developers are conscious of the importance of paying off technical debt items.

\begin{table}[th]
\caption{Amount of technical debt that was repaid.}
\label{tb:repayment_statistics}
\begin{center}
\resizebox{0.95\columnwidth}{!}{
\def\arraystretch{1.2}
\begin{tabular}{c|c|c|c|c}
\hline
\textbf{Project} & \textbf{\# Identified} & \textbf{\# Repaid} & \textbf{\% Repaid} & \textbf{\% Remaining} \\
\hline
Hadoop & 101 & 72 & 71.3 & 28.7 \\
\hline
Camel & 51 & 37 & 72.5 & 27.5 \\
\hline
Total & 152 & 109 & 71.7 & 28.3 \\
\hline
\end{tabular}
}
\end{center}
\vspace{-2mm}
\end{table}

\subsubsection{\textbf{(RQ3.2) Who repays technical debt?}}
\label{sec:repaid_person}

\begin{table}[b]
\caption{Who repaid technical debt.}
\label{tb:repayment_person}
\begin{center}
\resizebox{0.95\columnwidth}{!}{
\def\arraystretch{1.2}
\begin{tabular}{C{1.2cm}|C{1.2cm}|C{0.5cm}|C{0.5cm}|C{0.5cm}|C{0.5cm}|C{0.5cm}|C{0.5cm}}
\hline
\multirow{3}{*}{\textbf{Project}} & \multirow{3}{*}{\textbf{\# Repaid}} & \multicolumn{6}{c}{\textbf{\# Repaid by}} \\
\cline{3-8}
 & & \multicolumn{2}{c|}{\textbf{Creators}} & \multicolumn{2}{c|}{\textbf{Identifiers}} & \multicolumn{2}{c}{\textbf{Others}} \\
\cline{3-8}
 & & \#\footnotemark[\value{footnote}] & \% & \#\footnotemark[\value{footnote}] & \% & \#\footnotemark[\value{footnote}] & \% \\
\hline
Hadoop & 72 & 36 & 50.0 & 33 & 45.8 & 3 & 4.2 \\
\hline
Camel & 37 & 12 & 32.4 & 19 & 51.4 & 6 & 16.2 \\
\hline
Total & 109 & 48 & 44.0 & 52 & 47.7 & 9 & 8.3 \\
\hline
\end{tabular}
}
\end{center}
\vspace{-5mm}
\end{table}

As shown in Table~\ref{tb:repayment_person}, we distinguish between developers who create technical debt, those who identify it and other developers who participate in resolving it. 
We can see that most of the technical debt is repaid by those who identified it (i.e. 47.7\%), and those who created it (i.e. 44.0\%); while only 8.3\% debt is resolved by other developers.
This shows that developers take the responsibility to pay off most of the technical debt they identified or created themselves.

\subsubsection{\textbf{(RQ3.3) How long does it take to fix technical debt?}}

\begin{figure}[thpb]
  \centering
  \includegraphics[width=\linewidth, trim=0cm 0.0cm 0cm 0.5cm,clip=true]{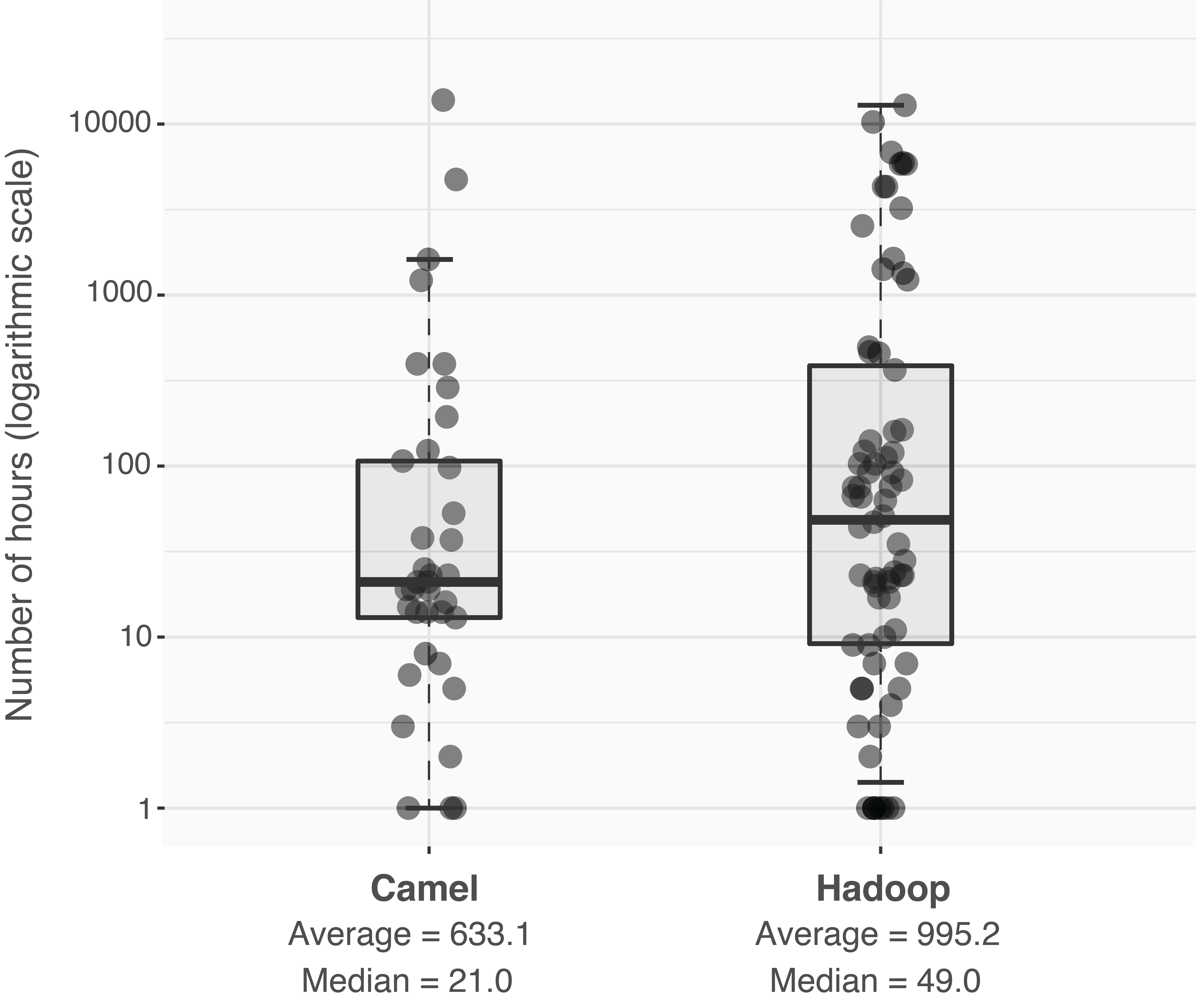}
  \caption{The time distribution of technical debt repayment in issue trackers.}
  \label{f:repayment_time}
\end{figure}

Fig.~\ref{f:repayment_time} shows the mean times, the median times, and the time distributions of technical debt repayment for the two projects.
With a visual inspection, we see that the time spent to fix technical debt in Hadoop and Camel varies.
We also observe that after the technical debt is reported (point zero in the y axis), most of the fixes happened in a short time compared to the average (67.0\% of the debt is repaid in the first 100 hours).

\begin{table}[b]
\caption{Repayment time comparison between different developers} 
\label{tb:repayment_test}
\begin{center}
\resizebox{\columnwidth}{!}{
\def\arraystretch{1.2}
\begin{tabular}{C{1cm}|C{2.35cm}|C{2.35cm}|C{0.9cm}|C{2.2cm}}
\hline
\multirow{2}{*}{\textbf{Project}} & \multicolumn{2}{c|}{\textbf{Average time spent on debt repayment (h)}} & \multirow{2}{*}{\textbf{\textit{p}-value}} & \multirow{2}{*}{\textbf{Cliff's delta}} \\
\cline{2-3}
 & \textbf{Creators} & \textbf{Identifiers} & & \\
\hline
Hadoop & 128.0 & 1510.8 & 0.031 & -0.303 (small) \\
\hline
Camel & 174.5 & 142.3 & 0.935 & 0.021 (negligible) \\
\hline
 & \textbf{Creators} & \textbf{Others} & & \\
\hline
Hadoop & 128.0 & 5730.3 & 0.028 & -0.777 (large) \\
\hline
Camel & 174.5 & 3104.3 & 0.851 & -0.069 (negligible) \\
\hline
 & \textbf{Identifiers} & \textbf{Others} & & \\
\hline
Hadoop & 1510.8 & 5730.3 & 0.080 & -0.626 (large) \\
\hline
Camel & 142.3 & 3104.3 & 0.463 & -0.210 (small) \\
\hline
\end{tabular}
}
\end{center}
\vspace{2mm}
\end{table}

Furthermore, we compare the time spent on resolving technical debt by different developers (Creators, Identifiers, and Others as discussed in Section~\ref{sec:repaid_person}).
More specifically, we compare repayment time distributions between pairs of developers (e.g. between creators and identifiers) using the Mann-Whitney test \cite{mann1947test} and Cliff's delta \cite{grissom2005effect} to determine the significance level and the effect size of the differences.
The result is demonstrated in Table~\ref{tb:repayment_test}.
There are notable differences between Hadoop and Camel.
In Hadoop, the repayment time of identifiers and others is longer than creators with statistical significance (\textit{p}-values are 0.031 and 0.028 respectively).
Moreover, the time difference between identifiers and others is at the margin of statistical significance (\textit{p}-value is 0.080).
According to the effect size, we observe that the difference between creators and identifiers is small, while the difference between identifiers and others is large.
Thus, technical debt in Hadoop is paid back the quickest by creators, followed with a small margin by identifiers, followed with a large margin by others.
In Camel, the situation is different as none of the time differences is statistically significant.
We only observe that the repayment time of others is much longer (on average) than creators and identifiers.

\begin{framed}
\noindent \textit{Most of the identified technical debt in issue trackers is resolved  (on average 71.7\%).
Debt identifiers and creators pay off most of the technical debt (47.7\% and 44.0\% respectively).
The median time and average time to repay technical debt are 25.0 and 872.3 hours.
In Hadoop, technical debt creators resolve it quicker than those who identify it or others.
}
\end{framed}

\section{Discussion}
\label{sec:discussion}

\textbf{Various types of technical debt are detected in issues, and they are complementary to those detected in source code comments.}
Comparing the types of technical debt we identified in issues (RQ1) against those types detected in source code comments by Potdat and Shihab \cite{maldonado2015detecting}, we find requirement, defect, design, test, and documentation debt appearing in both. 
However, although documentation and test debt are among the three most common types in issues, they are the two least common types in source code comments.
Meanwhile, design debt is the most common type in source code comments, but it is rather uncommon in issues.
Finally, code, build, and architecture debt are only detected within issues.
This means that the types of technical debt detected through issue trackers and source code comments have some overlap but they are also sufficiently distinct. Thus,  using each source (issues or source code comments) has its strengths and weaknesses.
Therefore, we argue that the two sources are complementary in detecting different types of technical debt. 
Researchers should take both sources into account to increase the completeness and accuracy of their detection tools.

\textbf{Approaches are required to identify technical debt in all three different cases (existing debt, during code review or after committing a patch)}.
Researchers should customize the identification approaches according to the characteristics of each case (see results of RQ1). 
For example the approach proposed by Dai and Kruchten \cite{dai2017detecting} can work for the first case but not for the other two cases.
Furthermore, the findings show that only 13.1\% of technical debt is reported by those that created it. 
We suggest that researchers look into this phenomenon and interview practitioners to find out why they usually do not report their own debt. 
Furthermore, we advise practitioners who deliberately incur technical debt, to report it in the issue tracker. 
This would accelerate the repayment of these debt items, even if that is performed by other developers. 

\textbf{Technical debt admitted in issues is resolved faster than in source code comments.}
Considering the results obtained from RQ3 in comparison with the study of Maldonado \textit{et al.} \cite{maldonado2017empirical}, we find that most of the technical debt is repaid or removed (71.7\% for debt within issues and 76.7\% for debt in code comments). Furthermore, a great percentage of technical debt is repaid or removed by debt creators (44.0\% for issues and 54.4\% for source code comments).
This indicates that developers consistently take care of SATD in both issues and source code comments, and debt creators often take the responsibility to resolve it.

Moreover, in Hadoop, it is noteworthy that debt creators repaid technical debt the fastest, followed by identifiers, and other developers. 
This is consistent with the intuition that certain people are better able to resolve TD depending on their familiarity with the problem at hand; creators being the most familiar, followed by debt identifiers, and others.
Therefore, we suggest that, in order to pay off TD faster, the repayment task should be assigned to debt creators.
In addition, comparing the TD repayment between issues and source code comments \cite{maldonado2017empirical}, we observe that debt within issues is resolved much quicker than in comments (i.e. for Hadoop, median of 2.0 days versus 159.0 days; for Camel, 0.9 days versus 18.2 days).
Therefore, we suggest that developers report TD that needs to be resolved immediately in issue trackers instead of commenting it in the source code.

\section{Threats to Validity}
\label{sec:validity}

\noindent \textbf{Threats to Construct Validity} concern the correctness of operational measures for the studied subjects.
Since only a small subset of issues in issue trackers contain technical debt statements \cite{bellomo2016got}, the sample (500 analyzed issues) may not represent the population (issues containing technical debt in general).
To minimize this threat, the analyzed sample was obtained randomly from all collected issues.

\noindent \textbf{Threats to Reliability} concern potential bias from the researchers in data collection or data analysis.
Since issues are written in natural language, they were identified and categorized manually.
To counter the threat of researchers biasing the manual analysis, the first and second author annotated the issue sample independently, and then discussed any differences to reach consensus on the classification.
Additionally, the level of agreement (Cohen's kappa) was 0.757, which indicates high inter-rater agreement. 
Finally, as aforementioned all data are publicly available in the replication package.

\noindent \textbf{Threats to External Validity} concern the generalization of findings.
In this study, we analyzed issues from two large open source projects, which both use JIRA as the issue tracker.
Thus, our findings may be generalized to other open source Java projects of similar size and complexity that use JIRA; we can not claim any further generalization.

\section{Conclusion}
\label{sec:conclusion}

In this paper, we explored SATD in issue trackers.
We found eight types of technical debt: architecture, build, code, defect, design, documentation, requirement, and test debt.
Code, documentation, and test debt are the three most common technical debt found in issue trackers. 
Furthermore, there are three cases of identifying technical debt in issue trackers: discovering existing debt and creating an issue for it, identifying debt in a patch during code review, or after the patch is committed in the system. 
Most of the technical debt is identified in the first and second cases. 
Only 13.1\% of technical debt is reported by debt creators.

For technical debt repayment, we found that on average 71.7\% of identified debt is repaid, and most of it is paid by debt identifiers and creators (i.e. 47.7\% and 44.0\%).
The median time and average time of debt repayment are 25.0 and 872.3 hours respectively.
Our results show that in Hadoop, the repayment time by creators is statistically significantly shorter than that of identifiers and others.
However, in Camel, we did not observe statistically significant differences between different types.


\bibliography{bibliography}
\bibliographystyle{IEEEtran}

\end{document}